\documentclass[runningheads]{llncs}
\usepackage[T1]{fontenc}
\usepackage{graphicx}
\usepackage{booktabs}
\usepackage{amsmath}
\usepackage{hyphenat}
\usepackage{adjustbox}
\usepackage{multirow} %for multirow
\usepackage{subfig}
\usepackage{caption}

\usepackage{float}
\usepackage{booktabs}
\usepackage{mwe}

\usepackage{algorithm,algpseudocode}
\usepackage[title]{appendix}%
\usepackage[misc]{ifsym}

\usepackage{mwe}
\begin{document}

\title{Improving Novel Anomaly Detection with
Domain-Invariant Latent Representations}

\titlerunning{Improving Novel Anomaly Detection by Learning
Domain-Invariant Representations in Latent Space}
% If the full title of your paper is short enough to also fit in the running head, you can omit the abbreviated paper title here. You can check as follows: if you comment out the \titlerunning line, something will appear in the header of all odd-numbered pages of your PDF from page 3 onward. This something is either the full title (in which case all is well), or the error message "Title Suppressed Due to Excessive Length". If this error message appears, you're going to want to provide an abbreviated title within the \titlerunning command, because if you won't do it, Springer will do it for you.

%N.B.: Author information (both in the \author{} and \authorrunning{} command) should only be present in the Camera-Ready Version of your paper. The version that you initially submit for review, ought to be double-blind. So, when initially submitting your paper, use:
%\author{Author information scrubbed for double-blind reviewing}

\author{Padmaksha Roy\inst{1} \orcidID{0000-0002-9571-1117}(\Letter) \and
Ming Jin\inst{1}\orcidID{0000-0001-7909-4545} \and Himanshu Singhal\inst{1} \orcidID{0000-0002-0474-8126} \and
Tyler Cody\inst{1} \orcidID{0000-0001-9215-5816} \and
Kevin Choi\inst{2} \orcidID{0009-0004-1818-2401} }
% You may leave out the orcidID information, if you want to.
% Use \corr to indicate the corresponding author. Note the spacing around the \corr command. Only one author can be the corresponding author.

%N.B.: comment out the \authorrunning{} command for the double-blind version of your paper submitted for review. Later, if your paper is accepted, use the command for the Camera-Ready Version.
\authorrunning{padmaksha roy et al.}
% First names are abbreviated in the running head.
% If there is one author, write 'A.L. Benjamin'.
% If there are two authors, write 'A.L. Benjamin and C.C. Broadus Jr.'
% If there are more than two authors, '[...] et al.' is used.

\institute{Virginia Tech, Blacksburg, Virginia, USA \email{\{padmaksha,jinming,himanshusinghal,tcody\}@vt.edu}
\and
AI Center of Excellence, Deloitte, Mclean, Virginia, USA \email{kevchoi@deloitte.com}}

\maketitle              % typeset the header of the contribution

\begin{abstract}
Zero-day anomaly detection is critical in industrial applications where novel, unforeseen threats can compromise system integrity and safety. Traditional detection systems often fail to identify these unseen anomalies due to their reliance on in-distribution data. Domain generalization addresses this gap by leveraging knowledge from multiple known domains to detect out-of-distribution events. In this work, we introduce a multi-task representation learning technique that fuses information across related domains into a unified latent space. By jointly optimizing classification, reconstruction, and mutual information regularization losses, our method learns a minimal(bottleneck), domain-invariant representation that discards spurious correlations. This latent space decorrelation enhances generalization, enabling the detection of anomalies in unseen domains. Our experimental results demonstrate significant improvements in zero-day or novel anomaly detection across diverse anomaly detection datasets.

\keywords{Representation Learning  \and OOD Detection \and Multi-task Learning.}
\end{abstract}

\section{Introduction}
Anomaly detection is a fundamental task in various applications, enabling the early identification of unusual patterns in network traffic, system logs, or user behavior that may signal intrusions or malicious activities \cite{zhou2022domain,wang2022generalizing}. As cyber threats evolve and novel attacks—such as zero-day vulnerabilities—emerge, traditional defenses often fall short, leading to severe disruptions and data breaches. In many real-world applications, training and test data stem from different distributions, making out-of-distribution (OOD) generalization a critical challenge. Standard deep neural networks excel when the training and testing data are drawn from the same distribution; however, their performance degrades when confronted with unseen domains. Existing approaches such as few-shot learning and meta-learning \cite{vuorio2019multimodal,sung2018learning,snell2017prototypical,vinyals2016matching,finn2017model} attempt to bridge this gap but often require target domain data during training or otherwise risk embedding biases from specific domains. Our approach addresses these challenges by targeting a latent space that embodies a minimal sufficient representation for the downstream task of OOD classification. We consider a scenario where the samples from different domains or datasets have distinct feature correlation structures. High-dimensional data poses unique challenges due to the curse of dimensionality. In such spaces, conventional distance measures lose their discriminative power because the relative contrast between the nearest and farthest neighbors diminishes—a phenomenon highlighted by the principle of concentration of distance.
Inspired by the principle of relevant information(PRI) preservation \cite{tishby2000information}, {we design a latent space classification loss that aims to regularize the latent space by minimizing the mutual information content between the  input and latent space, effectively decorrelating class-specific feature correlation information of the original data. To guarantee that the latent space preserves sufficient input information, we incorporate a reconstruction loss that compels the model to accurately reconstruct the input data from its latent embedding. This prevents over-compression and ensures that the latent space retains the necessary structure for the task. These two losses guide the cross-entropy loss to preserve only the relevant information required for accurate classification. Multi-task learning facilitates learning representations from multiple diverse domains and the joint optimization help improve generalization to unseen domains. By integrating these objectives, our framework works as a zero-shot multi-task learning system. We mix data from multiple source domains with cross-domain samples and also attempt to decorrelate dataset specific spurious correlation information with the mutual information (MI) penalty. This strategy ensures that the learned latent space is invariant to domain-specific correlation information, thereby enhancing generalization to unseen OOD classes without requiring any target domain data during training. Our main contributions can be summarized as follows:
\begin{itemize}
    \item We propose a novel classification framework that leverages mutual information regularization and reconstruction loss to guide the latent space toward retaining only the most relevant features for out-of-distribution (OOD) classification. The result is a \textit{compressed, invariant} representation that effectively discards \textit{spurious} domain-specific information.

    \item  We demonstrate that integrating data from multiple sources and cross-domains with varying \textit{correlation} patterns enhances \textit{coverage}, improving generalization to unseen domains. 

    \item  Our domain-invariant latent space analysis mitigates the adverse effects of \textit{high-dimensionality}. Experiments demonstrate an 8\%-15\% increase in average precision, and recall and a 4\%-9\% improvement in average AUC-ROC across all source/IN, cross-domain, and OOD datasets.
\end{itemize}

\section{Related Work}
Domain generalization techniques can be grouped into the following primary categories: domain invariant representation learning, meta-learning,latent dimension regularization, and metric learning.
\textit{1) Domain Invariant Representation Learning:} This method aims to identify domain invariant representations that can be extended to unseen domains. The crux of these strategies, as seen in works such as \cite{seo2020learning}, is to filter out domain-specific insights while maintaining cross-domain information. Notable studies employing autoencoders, such as \cite{ghifary2015domain}, amalgamate multiple domains during training, augmented by data enhancement techniques, to extract domain-invariant characteristics. These features then demonstrate superior generalization to out-of-distribution data. Another study, Maximum Mean Discrepancy Adversarial Autoencoder (MMD-AAE) \cite{li2018domain}, in the context of few-shot learning, emphasizes aligning varied domain distributions to a generic prior distribution while engaging in adversarial feature learning. An innovative approach is suggested in \cite{chattopadhyay2020learning}, where a domain-centric masking technique is applied to learn both domain-specific and domain-invariant features. This will facilitate efficient source domain classification and sufficient generalization to target domains. In \cite{liang2021boosting}, a noise-enhanced supervised autoencoder reconstructs and classifies both inputs and their reconstructions, using intra-class correlation to show improved feature discrimination and generalization. Moreover, the authors \cite{jin2020feature} propose domain generalization through domain-invariant representation that uniformly distributes across multiple source domains. Their approach employs moment alignment of distributions and enforces feature disentanglement via an entropy loss. The DIFEX \cite{lu2022domain} paper employs knowledge distillation to capture internally-invariant Fourier phase features and aligns cross-domain correlations to extract mutually-invariant representations.

\textit{2) Meta-learning:} This approach employs learning from several related tasks for domain generalization, as observed in works such as \cite{finn2017model,li2018deep,li2018learning}. The study in \cite{erfani2016robust} introduces a technique to discern a domain interdependent projection leading to a latent space. This space minimizes biases in the data while preserving the inherent relationship across multiple domains. Model Agnostic Meta-Learning (MAML) has also been extended to latent dimension settings by performing the gradient-based adaptation in the low dimensional space instead of the higher dimensional space of model parameters \cite{rusu2018meta}. Zero-shot learning \cite{wang2019survey} aims at learning models from seen classes and inferring on samples whose categories were unseen during the training process.

\textit{3) Information Bottleneck Principle and Metric Learning:} In contrast to the aforementioned methodologies, our strategy propels direct disentanglement or decorrelation between multiple training domains. 
% Decorrelating spurious feature correlations to isolate signals from background noise has been successfully used in the physics and statistics community before. 
An information-theoretic perspective on variance-invariance-covariance has been provided here \cite{shwartz2023information} in the context of self-supervised learning which helps to achieve generalization guarantees for downstream supervised learning tasks. Adversarial learning-based domain adaptation methods are prone to negative transfer which hurts the generalization performance
\cite{jeon2021mutual}. 
% In our method, class-invariant features are indirectly learned in the latent dimension by minimizing mutual information between latent and prior space.
% \textit{4) Metric Learning:}
% % Need to add some background here. The connection is too obvious to not mention unless there is a reason otherwise. I think metric learning focuses on adding distance metrics to the loss function, this work focuses on mutual information. And this work is different from metric learning methods that do incorporate distance because it is a multi-task, zero-shot classification method.
Metric learning aims to learn a representation function that can map higher-dimensional data to a latent embedded space. The authors \cite{venkataramanan2021takes} propose mixing target labels with training samples to improve the quality of representations or embeddings for classification purposes. 
% In deep metric learning, we try to learn non-linear mapping from input space to low-dimensional latent space \cite{oh2016deep}. Various mix-up techniques have been recently proposed to improve the quality of representation learned by adopting some interpolation techniques between pairs of input samples and their labels \cite{zheng2019hardness}. 

% On the other hand, we follow a multi-task representation learning approach that aims to learn a robust feature representation in the latent space by leveraging multiple sources and cross-domain feature information.

\textit{4) Other Related Works:} The authors \cite{hendrycks2016baseline} suggest using the statistics of softmax outputs to estimate both the probability of error and the likelihood of a test sample being out-of-domain. They compare the performance of this approach by directly using the raw softmax output probabilities as a measure of confidence. The paper \cite{xu2018deep} addresses the problem of domain shift when a learned model tends to degrade heavily on a target domain via unsupervised domain adaptation by learning a common feature map from multiple source domains by minimizing the domain distribution discrepancy between those multiple source domains.
% This work \cite{yang2022openood} acknowledges the fact that Out-of-distribution (OOD) detection is critical for safety-critical applications but lacks a unified benchmark, leading to inconsistent comparisons, prompting the development of the OpenOOD codebase to provide a comprehensive benchmark. They also emphasize how OOD detection is closely linked to related fields such as anomaly detection (AD), open set recognition (OSR), and model uncertainty, as methods created for one area are often applicable to others. 
The authors in \cite{sun2022out} use nearest-neighbor distance for flexible OOD detection without strict assumptions, while \cite{arjovsky2019invariant} address spurious correlations by developing causal tools to distinguish invariant features, thereby improving generalization.

Our approach is inspired by the principle of relevant information (PRI) \cite{tishby2000information}, aiming to learn a compressed latent space that retains only the relevant information for downstream tasks. By combining data from multiple domains and de-correlating their spurious correlations, we encourage the network to learn invariant representations. This multi-task representation learning method ensures that the latent space captures minimal, sufficient information for classification while discarding irrelevant, domain-specific details.

\section{Problem Formulation}
In our domain generalization problem, let $\mathcal{C} = \{0,1,\ldots,K\}$ denote the complete set of class labels, which we partition into three disjoint subsets;
$
\mathcal{C} = \mathcal{C}_s \cup \mathcal{C}_c \cup \mathcal{C}_o,\quad \mathcal{C}_s \cap \mathcal{C}_c = \mathcal{C}_s \cap \mathcal{C}_o = \mathcal{C}_c \cap \mathcal{C}_o = \emptyset.
$
For example, if $\mathcal{C}_s = \{1,3,5\}$ (source domain), then $\mathcal{C}_c = \{2,4,6\}$ (cross-domains) and $\mathcal{C}_o = \{9,10,11\}$ (OOD). During training, we have access only to samples from the source and cross-domains. Formally, the training set is defined as
$
\mathcal{S}_{\text{train}} = \bigcup_{i=1}^{M} \{(x_j^i, y_j^i) \mid y_j^i \in \mathcal{C}_s \cup \mathcal{C}_c,\; j=1,\ldots,N_i\},
$
where $M$ is the total number of tasks and $N_i$ is the number of samples in task $i$. Our objective is to learn a model that generalizes to unseen OOD classes $y \in \mathcal{C}_o$ by leveraging multi-task representation learning. We enforce domain-invariant feature extraction through joint optimization over classification, reconstruction, and mutual information regularization losses, thereby encouraging a disentangled latent space that tends to forget spurious correlations. 
\begin{figure*}
\small
   \centering   \includegraphics[width=12cm,height=7cm]{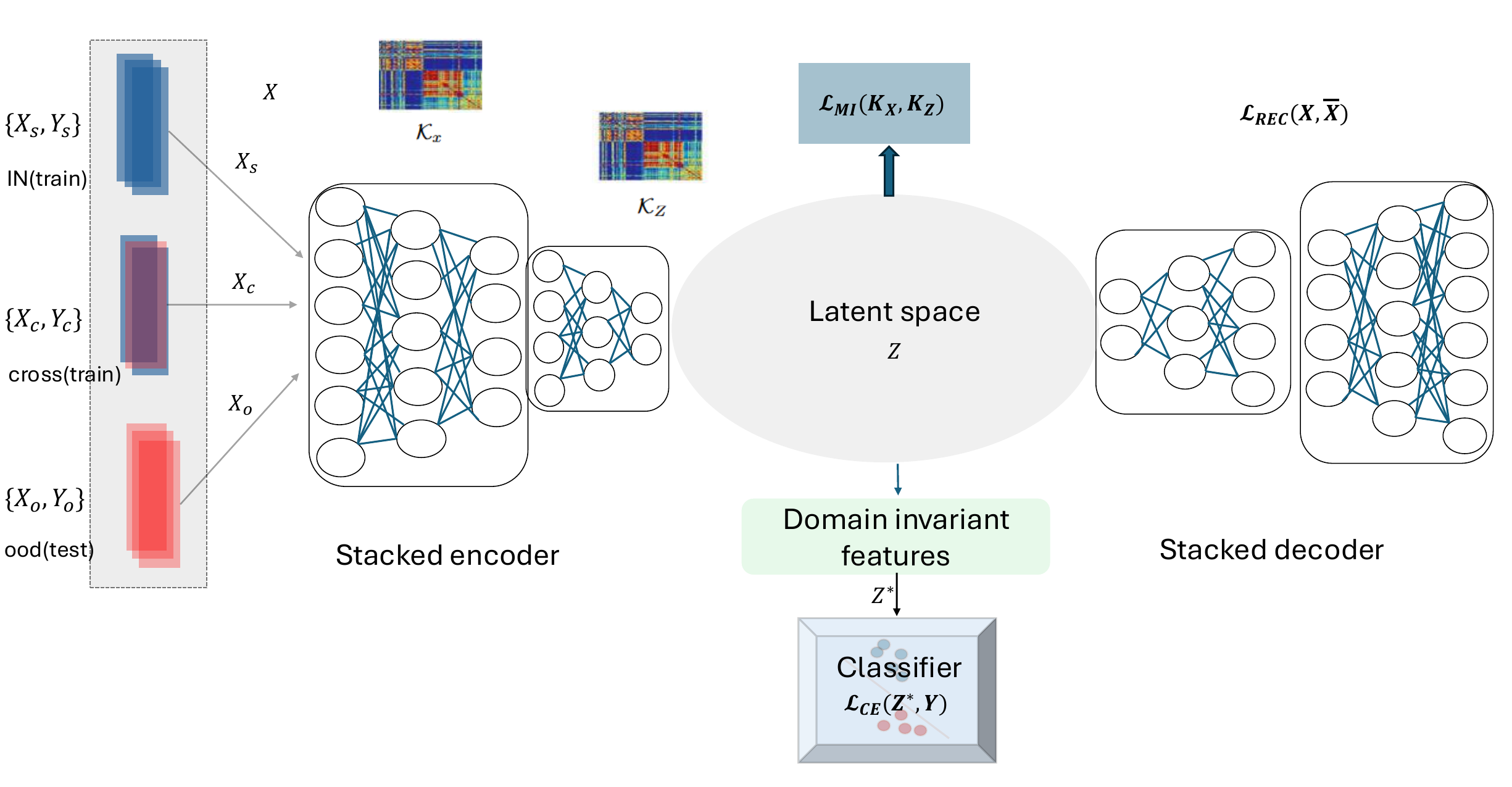}
   \caption{Training the Multi-task Latent Space Regularized Encoder-Decoder Model (MTLS-RED). During testing, the trained latent space is directly used to classify new samples.}
   \label{fig: DLSCA}
\end{figure*}
The extension to multiple domains necessitates the definition of a multi-task learning objective over all the M source and cross domains which can be given as
\begin{equation}
\begin{split}
    \mathcal{L}_{rec}\left( {S}_{train};\theta, \phi  \right) =   \sum_{i =1}^M \Big \|f_\theta^{(i)} \left(g_\phi^{(i)} \left(X^{i}\right) \right)- X^{i} \Big \|^2_2 \\
    % +  \sum_{\substack{i,j =0 \\j \neq M-i;}}^{N } \Big \| \left(g_\phi^{(i)} \left(X^{i}\right) \right)- 
    % \left(g_\phi^{(M-j)} \left(X\right) ^{M-j}\right) \Big \|^2_{2} 
\end{split}
\end{equation}
In this expression, $g_\phi^{(i)}$ and $f_\theta^{(i)}$ denote the encoder and decoder functions respectively for each of the $M$ sources and cross domains, $X^i$ is the input training data from a particular domain. We aggregate the reconstruction loss across different source and cross-domain datasets, ensuring that the total loss accounts for all input domains. Basically, the reconstruction error is computed separately for each domain and then summed to form the overall reconstruction loss.

\subsection{Mutual Invariance Regularization}
% To learn a domain-invariant latent space, we try to de-correlate the latent space from the input space in their kernel space (mutual information minimization) between the prior and the latent space which acts as a regularization technique. 
% \subsubsection{Kernel-based Renyi's Entropy and Joint Entropy to measure Mutual Information between Input and Latent Space}
In information theory, the dependence measure or the total correlation between the feature variables is measured as the statistical independence in each dimension and is expressed as the 
Kullback Leibler(KL) divergence between the joint probability distribution and the marginal distribution of the features \cite{yu2021measuring}. 
% The total correlation($\rho$)  can be estimated as
% \begin{equation}
%     \begin{split}
%         \hat{\rho}(X^{1},X^2,...,X^M ) &= D_{KL} \Bigl \{Pr \left(X^{1},X^2...,X^M \right) || \\ \prod_{i=1}^M Pr\left(X^i \right) \Bigr\}  
%          &=  \left[  \sum_{i=1}^{M} H \left(X^i \right) \right] - H \left(X^1,X^2...,X^M \right) .  
%     \end{split}
% \end{equation}
We enforce de-correlation between the input and the latent kernel space—spanning multiple source and cross-domains—by introducing a mutual information minimization penalty that explicitly reduces dependencies between input and latent space kernels in the form of decorrelation.
% This promotes a disentangled latent space, effectively eliminating spurious correlations inherent in the high-dimensional input domains.
The matrix-based Renyi's second-order entropy \cite{yu2021measuring} of a normalized positive definite(NPD) matrix $\mathcal{K}_x$, estimated on $l \times l$ samples in the input space, where $l$ is the batch size, can be given as 
 \begin{equation}
 \begin{split}
       \hat{H_2}(\mathcal{K}_x)
    = \frac{1}{1-\alpha}\log_{2} \left(\sum_{k=1}^l \lambda_{k}(\mathcal{K}_x)^\alpha \right),
\end{split}
\end{equation}    
where the Gram matrix $\mathcal{K}_x$ is obtained by evaluating the positive definite (PSD) kernel on all $l$ pairs of training samples in a batch of training data, that is, 
% $\left(\mathcal{K}_x\right)_{i,j}  = \mathcal{K}_x\left( x_{i}, x_{j}\right)$ 
and $\lambda_{k}({X})$ denotes the $k^{th}$ eigenvalue of the input kernel matrix $\mathcal{K}_x$ of the $l_{th}$ batch, 
% $\mathcal{K}_{i,j} = \frac{1}{l} \frac{\mathcal{K}_{i,j}}{\sqrt{\mathcal{K}_{i,i} \mathcal{K}_{j,j}}}
Here, $\alpha = 2$ considering Renyi's second-order entropy.
% and $\mathcal{G}$ represents a kernel Gram matrix obtained from evaluating a positive definite kernel on all pairs of samples.
% i.e., $\mathcal{G}_{ij} = \mathcal{K}\left( x_i, x_j\right)$ 

% \begin{equation}\label{entropy_prior}
% \hat{H}_{2}(X) = -\log \left( \frac{1}{N^2} \sum_{i=1}^{N} \sum_{j=1}^N \mathcal{G}_{{\sigma}_{\sqrt{2}}}(x_{i} -x_{j})\right).
% \end{equation}
Similarly, Renyi's quadratic entropy of the latent space kernel $\mathcal{K}_Z$ of size $l \times l$ is estimated as
 \begin{equation}
 \begin{split}
       \hat{H_2}(\mathcal{K}_z) 
    = \frac{1}{1-\alpha}\log_{2} \left(\sum_{k=1}^l \lambda_{k}(\mathcal{K}_z)^\alpha \right),
\end{split}
\end{equation}    
The argument in equation (3) is called the information potential.
In the above section, we use the matrix-based second-order Renyi's entropy $\left(\alpha =2 \right)$  \cite{yu2021measuring} to evaluate the entropy or the uncertainty of the latent and the input space in terms of the normalized eigenspectrum of the Hermitian matrix of the projected data in the Hilbert space. Now, we can estimate the matrix-based second-order joint entropy between the latent space kernel $Z$ and the input space kernel $X$  as
\begin{equation}
\hat{H_{2}}\left(\mathcal{K}_x,\mathcal{K}_z\right) = H_{2} \left(\frac{\mathcal{K}_x \circ \mathcal{K}_z} {tr\left( \mathcal{K}_x \circ \mathcal{K}_z \right)} \right),
\end{equation}
where $\circ$ represents the Hadamard product.
Based on the above definitions, we calculate the joint entropy of the latent and the input space with the help of the matrix-based normalized Renyi's entropy of the latent space and the input space kernels. The joint entropy is used to derive the mutual information between the input and the latent space.
\subsubsection{The Mutual Information Divergence}  We use the matrix-based mutual information divergence to estimate the mutual information between the latent and input space kernels. Minimizing the mutual information indirectly results in de-correlating the feature correlation that exists in the original input space which helps in improving the generalization performance. The mutual information during each batch of the training can be estimated as
\begin{equation}\label{eq:MI}
\begin{split}   
    \hat{MI}(\mathcal{K}_x ; \mathcal{K}_z) & = 
      \hat{ H_2} \left(\mathcal{K}_x \right) + \hat {H_2} \left(\mathcal{K}_z \right) - \hat{H_2} \left(\mathcal{K}_x,\mathcal{K}_z \right),
\end{split}
\end{equation}
where $\hat{H_2} \left(\mathcal{K}_X,\mathcal{K}_Z \right),$ is the second-order joint entropy between the latent and the input kernel space. Minimizing this divergence as a regularization penalty in the final loss objective will aid in preserving useful disentangled information in the latent space during each iteration of the training process.
\subsection{The Multi-Task Learning Objective}
In our latent space multi-task learning approach, we leverage the label information of the multiple source and cross-domain encoded data in the latent space during the training process. In our approach, we do a joint optimization of the classification and the reconstruction loss along with the mutual information penalty in the latent space. The total loss calculated over all the $M$  tasks can be written as
\begin{equation}
\begin{split}
    \mathcal{L} \left(\mathcal{S}_{train},Z;  {\phi},{\theta}, \sigma \right) =  \min_{\phi,\theta,\sigma} 
    \sum_{i=1}^{M}\Bigr \{ \mathcal{L}_{ce}\left( g_{\phi} \left(X^i \right),y^i \right) \\ + \beta \cdot \mathcal{L}_{MI}\left( X^i;Z^i, \sigma \right) 
    + \lambda \cdot \mathcal{L}_{rec}\left(X^i;\phi, \theta \right)
    \Bigl \},
\end{split}
\end{equation}

% \begin{equation}
% \begin{split}
%     \mathcal{L} \Bigl(\mathcal{S}_{train}, Z; \, {\phi}, {\theta}, \sigma_x, \sigma_y \Bigr) 
%     \;=\; \min_{\phi, \theta, \sigma_x, \sigma_y} 
%     \frac{1}{M} \sum_{i=1}^{M} \Bigl\{ \,
%     \mathcal{L}_{ce}\!\Bigl( g_{\phi} \bigl(X^i \bigr),\, y^i \Bigr) 
%     \;+\; \beta \,\cdot\, \mathcal{L}_{MI}\!\Bigl( X^i;\, Z^i;\, \sigma_x,\sigma_y \Bigr) \\[6pt]
%     \;+\; \lambda \,\cdot\, \mathcal{L}_{rec}\!\Bigl(X^i;\,\phi, \theta \Bigr)
%     \Bigr\},
% \end{split}
% \label{eq:augmented_loss_kernel}
% \end{equation}
\begin{algorithm}[h!]
\caption{The Multi-task Latent Space Regularized Encoder-Decoder Model (MTLS-RED)}
\label{alg:mtls-red}
\begin{algorithmic}
\State \textbf{Input:}
\State Source domain data $\{X_s{_{1}},X_s{_{2}},...,X_s{_{m}}\}$, $X_s \in 
\mathbf{R}^d$, $\forall{m} \in \{1,2,3,..\}$ 
\State Cross-domain data $\{X_c{_{1}},X_c{_{2}},...,X_c{_{n}}\}$, $X_c \in \mathbf{R}^d$, $\forall{n} \in \{4,5,6,..,\}$ 
\State Out-of-distribution (OOD) datasets $\{X_{o_{1}}, X_{o_{2}}, ..., X_{o_{k}}\}$, $X_{o} \in \mathbf{R}^d$, $\forall{k} \in \{7,8,9,..,\}$ (used for testing only)
\State Source and cross-domain labels $\{y_i^m\}_{i=1}^n$, $\forall{m} \in \{1,2,3,4,5,6,...,M\}$
\State Initialize encoder (E) and decoder (D) weights: 
\State \quad $\mathbf{W}_\phi \in \mathbf{R}^{d_x \times d_z}$, $\mathbf{W}_\theta \in \mathbf{R}^{d_z \times d_x}$ 
\State Initialize kernel bandwidths: $\sigma_x$, $\sigma_y$ (learnable)
\State Set learning rates $\alpha_1, \alpha_2, \alpha_\sigma$

\While{not end of epochs}:
\For {batch = 1 to total batches $N$}:
    \State Sample mini-batch data $\{X_i\}_{1}^{l} \in \mathbf{R}^d $, where $l$ is batch-size
    \State Compute RBF kernels for input space $\mathcal{K}_{x_{l}}$ and latent space $\mathcal{K}_{z_{l}}$ of size $l \times l$
    \State Compute mutual information between input space $X_l$ and latent space $Z_l$ using matrix-based Rényi's entropy:
    \[
    MI\left(\mathcal{K}_{{X}_{l}};\mathcal{K}_{{Z}_{l}} \right)
    \]
    \State Perform a forward pass on encoder $E\left( X_{{\phi}_i}\right)$
    \State Compute total batch loss:
    \[
    \mathcal{L}_{l}= \mathcal{L}_{ce} \left({X^l,y^l}\right)+\lambda \mathcal{L}_{rec}\left(X^l, X^{{l}^\prime}\right)  + \beta \mathcal{L_{MI}} \left({X}^l || {{Z}^l} \right)
    \]
    \State Update $\mathbf{W}_\phi$, $\mathbf{W}_\theta$, and $\sigma_x, \sigma_z$
    \State $  \mathbf{W}_{\phi_{t+1}} \leftarrow \mathbf{W}_{\phi_{t}} - \alpha_1 \nabla_\phi \mathcal{L}_l\left (\theta,\phi, \sigma \right) $
    \State $  \mathbf{W}_{\theta_{t+1}} \leftarrow \mathbf{W}_{\theta_{t}} - \alpha_2 \nabla_\theta \mathcal{L}_l\left (\theta,\phi, \sigma\right) $
    \State $  \sigma_{x_{t+1}}, \sigma_{z_{t+1}} \leftarrow \sigma_{x_{t}}, \sigma_{z_{t}} - \alpha_\sigma \nabla_{\sigma} \mathcal{L}_l\left (\theta,\phi, \sigma \right) $
\EndFor
\EndWhile
\State \textbf{Output:} Trained MTLS-RED model with optimized encoder-decoder weights $\mathbf{W}_\phi, \mathbf{W}_\theta$ and learned kernel bandwidths $\sigma_x, \sigma_y$
\end{algorithmic}
\end{algorithm}
where, $\mathcal{L}_{ce}$ is the cross-entropy loss calculated on the latent space encoding considering  the binary classification problem, given as,
\begin{equation*} \mathcal{L}_{ce}\left( g_{\phi} \left(X^i\right),y^i \right) \\ = - \left(y^i \log \left(\mathcal{S}_y^i\left( g_{\phi} \left(X^i \right) \right)\right) \\ + \left( 1 - y^i \right) \log \left( 1 - \left( \mathcal{S}_y \left(g_{\phi} \left(X^i \right)\right)
    \right)\right)\right) 
\end{equation*}
$\mathcal{L}_{MI}$ is the disentanglement or de-correlation loss between the latent space and the input space expressed in the form of mutual information divergence measured in their kernel space, given in eq: \ref{eq:MI}, $\mathcal{S}_y$ is the softmax function applied on the encoded data $g_\phi(x)$, $\mathcal{L}_{rec}$ is the reconstruction loss, $\phi$, $\theta$ are the encoder and decoder parameters. The $\sigma$ represents the kernel bandwidth, a crucial parameter for estimating mutual information between the input and latent space.

We guide the cross-entropy loss by incorporating mutual information regularization between the latent and input spaces. This regularization discourages the retention of irrelevant information in the latent representation, with its strength governed by the hyperparameter $\beta$ The parameter 
$\beta$ regulates the trade-off between reducing dependencies in the latent space and maintaining classification performance. During joint optimization, we aim to balance the reconstruction loss and mutual information regularization. The parameter $\lambda$ controls the reconstruction weight, determining the extent of compression we want to enforce in the latent space.
\section{Experiments} In this section, we demonstrate the performance of our proposed model on benchmark cybersecurity and healthcare datasets.
\subsection{Dataset}
\begin{itemize}
  \item \textbf{ 
 % Communications Security Establishment (CSE) and the Canadian Institute for Cybersecurity (CIC) Intrusion Detection System Dataset
 CSE-CIC-IDS2018 \cite{CIC_datasets}} This is a publicly available cybersecurity dataset that is made available by the Canadian Cybersecurity Institute (CIC). It consists of 7 major kinds of intrusion datasets. We use SOLARIS, GOLDENEYE as source domain data, INFILTRATION, BOTNET as cross-domain data, and RARE, SLOWHTTPS, HOIC and a BENIGN dataset of a different day as the OOD test classes.
 \item  \textbf{CICIoT 2023 \cite{CIC_datasets}} This is a state-of-the-art dataset for profiling, behavioral analysis, and vulnerability testing of different IoT devices with different protocols from the network traffic, consisting of 7 major attack classes. We use BENIGN, DoS, and DDoS as source data, RECON,  as cross-domain data and WEB, MIRAI as OOD test data.
 \item  \textbf{CICIoMT 2024 \cite{CIC_datasets}} This is a benchmark dataset to enable the development and evaluation of Internet of Medical Things (IoMT) security solutions. The attacks are categorized into five classes. We use BENIGN, DDoS, DoS as source-domain, RECON, and SPOOFING  as cross-domain, and MQTT as OOD data.
 \item  \textbf{Arrythmia} This dataset is about atrial fibrillation (also called AFib or AF) which is a quivering or irregular heartbeat (arrhythmia) that can lead to blood clots, stroke, heart failure, and other heart-related complications. The dataset contains five classes/categories: N (Normal), S (Supraventricular ectopic beat), V (Ventricular ectopic beat), F (Fusion beat), and Q (Unknown beat). 
  % \item  \textbf{EMG Gesture Recognition Dataset} The EMG Signal for Gesture Recognition dataset is a widely used dataset in the field of bioinformatics and human-computer interaction, particularly for developing systems that recognize hand gestures based on electromyography (EMG) signals. Common gestures include a fist, open hand, pointing, and various finger movements.  
  % The dataset consists of EMG signals collected using sensors placed on the forearm. The dataset typically includes recordings of multiple hand gestures. 
 \end{itemize}

 \subsection{Baselines} We consider the following models related to multi-task representation learning and few-shot learning as baselines.
\begin{itemize}
    \item \textbf{Correlation Alignment for Deep Domain Adaptation (CORAL)} \cite{sun2016deep} This work has been employed for supervised domain adaptation, aligning source and target covariances to enhance OOD generalization.
    
    \item \textbf{Multi-task Autoencoder (MTAE)} \cite{ghifary2015domain} This encoder-decoder model optimizes reconstruction error across multiple domains in a supervised manner, jointly training sources and cross-domain data with label information in a two-stage process.
    \item \textbf{Minimum Mean Discrepancy-Autoencoder(MMD-AE)}\cite{li2018domain,sathya2022adversarially} This paper uses the MMD measure as regularization for domain generalization between multiple cross-domain data. We use it as a few-shot learning method where the cross-domain data are added to improve the OOD generalization.
    \item \textbf{Noise Enhanced Supervised Autoencoder (NSAE)} \cite{liang2021boosting}This model jointly predicts input labels, reconstructs inputs as noisy samples, and refines them through an additional fine-tuning step using a supervised classifier.     
    \item \textbf{Domain-invariant Feature Exploration for Domain Generalization (DIFEX) } \cite{lu2022domain} This paper utilizes mutual invariance to extract cross-domain features for OOD classification, capturing domain-specific semantics through internal invariance while preserving shared information, and extends CORAL with an additional regularization term.
\end{itemize}
\begin{figure*}[htp]
    \centering
   \includegraphics[width=\textwidth,height=1.8cm]{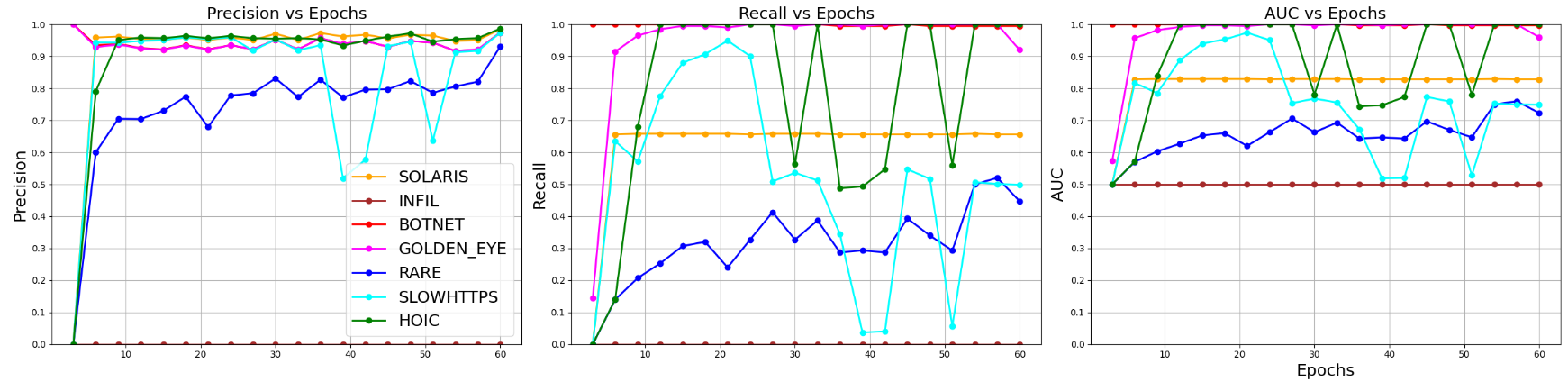} \\
    \makebox[\textwidth]{(a)} \\[10pt]

    \includegraphics[width=\textwidth,height=1.8cm]{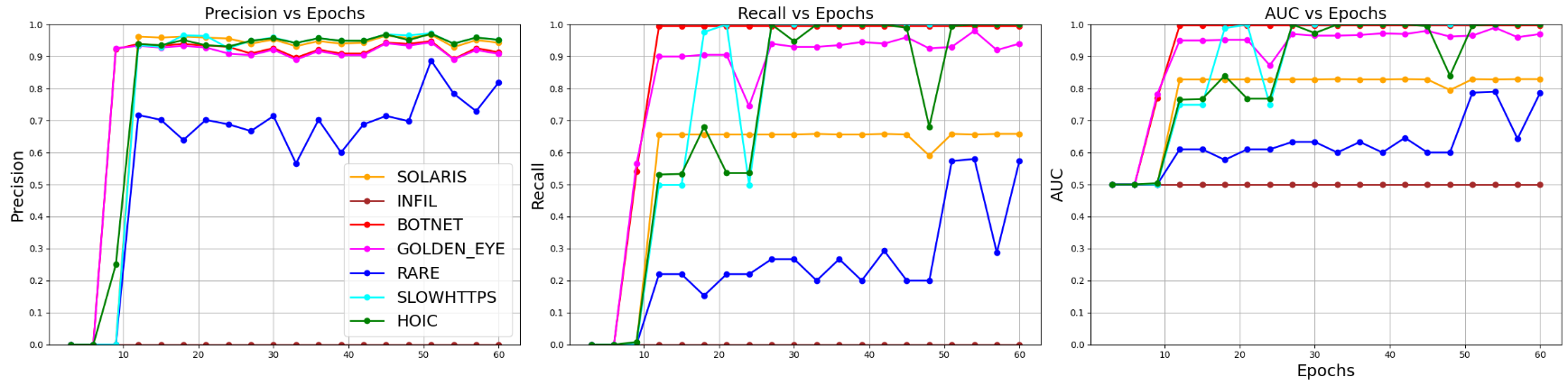} \\
    \makebox[\textwidth]{(b)} \\[10pt]

    \includegraphics[width=\textwidth,height=1.8cm]{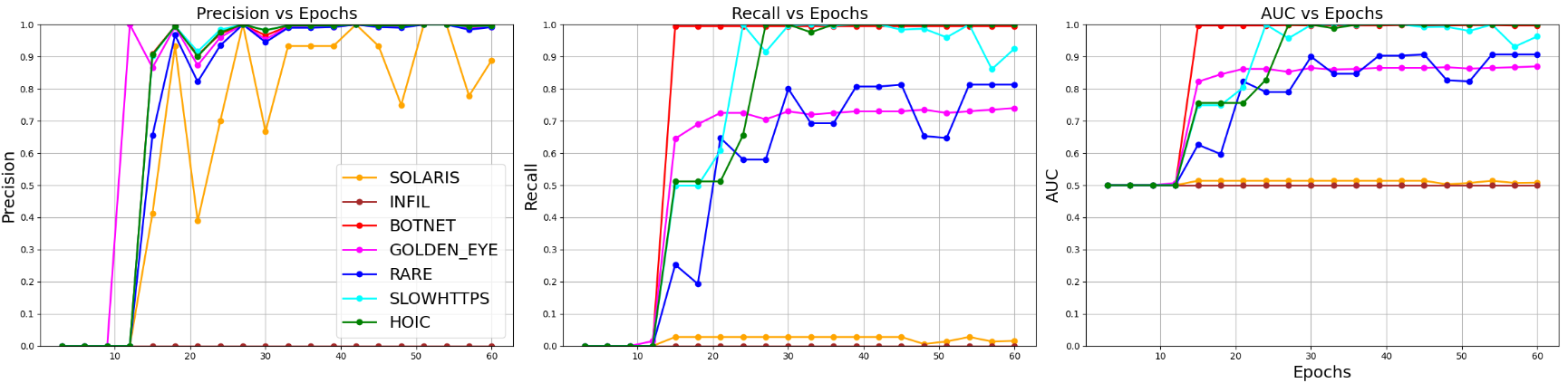} \\
    \makebox[\textwidth]{(c)} \\[10pt]

    \includegraphics[width=\textwidth,height=1.8cm]{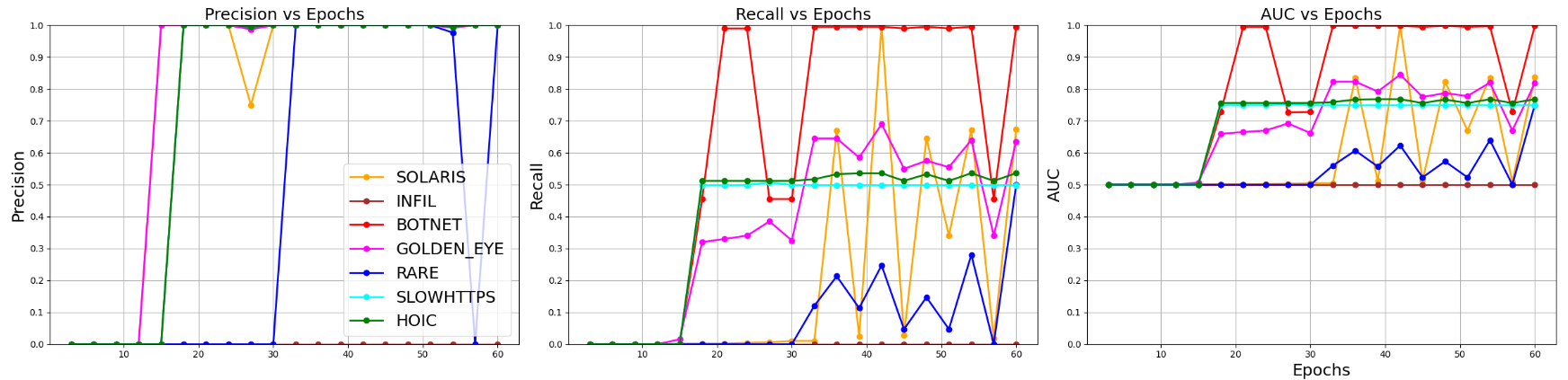} \\
    \makebox[\textwidth]{(d)}

    \caption{Precision, recall, and accuracy plots for the rarest class (RARE, in blue), which has only 525 samples in the CIC-IDS dataset using training data from GOLDENEYE (source) and BOTNET (cross) domains. Figures (a) show precision, recall, and AUC over epochs without regularization on validation data; (b) apply MI = 0.01, reconstruction = 0.99; (c) apply MI = 0.99, reconstruction = 0.01, (d)use equal weights of 0.5. High MI regularization (case (c)) leads to over 10-20\% improvement and stability across all metrics. Higher MI penalty helps in achieving better classification of the RARE class}
    \label{Fig:precision_recall_auc_rare}
\end{figure*}
\subsection{Training Strategy} To achieve robust generalization, we arbitrarily categorize the datasets into three groups: source domain datasets, cross-domain datasets, and out-of-distribution (OOD) datasets. The OOD datasets are reserved exclusively for testing purposes, serving as an evaluation benchmark for assessing model generalization. Our training strategy focuses on enhancing OOD performance by leveraging source domain data to improve learning on cross-domain datasets. To accomplish this, we systematically mix different proportions of source domain data with cross-domain data, integrating them into the benign dataset to construct the final training set. Additionally, we experiment with different combinations of source and cross-domain datasets to identify the most effective configurations for improving coverage across all three dataset categories—source, cross-domain, and OOD. Our training strategy is detailed in MTLS-RED Algorithm~\ref{alg:mtls-red}.

\subsubsection{Selecting the cross-domains, source and OOD domains}
In \cite{dong2022first}, the authors argue that learning a model that generalizes to unseen data can be facilitated when the \textit{covariance} (or correlation structure) among features is well-conditioned and sufficiently diverse. In other words, if the training data exhibit meaningful variations or “patterns of dependencies” across features, then a function that captures those variations can more reliably extrapolate beyond the training distribution. Hence, by adding source domain data (with one correlation structure) to cross-domain data (with a different correlation structure), we produce a more \emph{varied} training distribution—one that exposes the learner to multiple ways in which features can co-vary. The learner, in turn, is incentivized to find a representation that extracts the stable, non-spurious relationships across these distributions.

\begin{table*}[t]
\centering
\renewcommand{\arraystretch}{1.2}
\setlength{\tabcolsep}{5pt}
\caption{We report \textbf{accuracy} (with standard deviation) of the proposed and baseline methods on the CIC-CSE-IDS dataset, where cross-domain data is gradually added to the source domain during training in the range (0–50\%). 
% For instance, 0\% indicates no cross-domain data, while 50\% means half of the source anomalies are replaced with cross-domain anomalies, mixed with BENIGN samples. 
Best test accuracies for each model are highlighted. The OOD domains are used only for test/evaluation purposes. During train, each anomaly dataset has equal amount (50\%) of  BENIGN samples added to it, i.e, the train and test datasets are balanced (equal normal and anomaly samples).}
\label{Tab:1}
\resizebox{1\textwidth}{!}{
\begin{tabular}{lcccccccccc}
\toprule
{} & {} &  \multicolumn{2}{c}{\textbf{SOURCE DOMAIN}} & \multicolumn{2}{c}{\textbf{CROSS DOMAIN}} & \multicolumn{4}{c}{\textbf{OOD DOMAIN}} \\
\cmidrule(r){3-4} \cmidrule(r){5-6} \cmidrule(r){7-10}
\textbf{Model} & \textbf{Percent}  & SOLARIS & GOLDEYE & INFIL & BOTNET & RARE & HOIC & HTTPS & BENIGN  \\
\midrule

MTAE 
& $0\%$ & \textbf{99.97 (2.5)} & 92.70 (1.3) & 04.00 (0.1) & 09.50 (0.2) & 00.30 (0.0) & 00.38 (0.0) & 02.30 (0.1) & 97.98 (2.4) \\
& $20\%$ & 99.50 (2.5) & \textbf{99.70 (2.5)} & \textbf{69.40 (2.5)} & \textbf{79.50 (2.5)} & 61.30 (1.5) & 27.38 (0.7) & 32.30 (0.8) & \textbf{98.11 (2.5)} \\
& $30\%$ & 81.60 (2.0) & 79.60 (2.0) & 69.70 (1.7) & 78.50 (2.0) & 65.40 (1.6) & \textbf{50.00 (1.3)} & 42.40 (1.1) & 62.00 (1.6)  \\
& $50\%$ & 61.61 (1.5) & 61.30 (1.5) & 31.30 (0.8) & 61.60 (1.5) & \textbf{72.60 (1.8)} & 48.50 (1.2) & \textbf{69.80 (1.7)} & 83.32 (2.1) \\
\midrule

MMD-AE
& $0\%$ & \textbf{99.99 (2.5)} & \textbf{99.29 (2.5)} & 00.53 (0.0) & \textbf{99.98 (2.5)} & \textbf{55.47 (1.4)} & \textbf{99.98 (2.5)} & \textbf{99.69 (2.5)} & \textbf{99.71 (2.5)} \\
& $20\%$ & 99.98 (2.5) & 92.55 (0.1) & \textbf{22.55 (0.6)} & 99.83 (2.5) & 12.19 (0.3) & 41.34 (1.0) & 98.77 (2.5) & 98.67 (2.5) \\
& $30\%$ & 99.78 (2.5) & 02.52 (0.1) & 15.35 (0.4) & 99.83 (2.5) & 12.19 (0.3) & 41.27 (1.0) & 99.69 (2.5) & 98.61 (2.5)  \\
& $50\%$ & 99.96 (2.5) & 01.88 (0.1) & 12.30 (0.3) & 99.83 (2.5) & 14.13 (0.4) & 41.12 (1.0) & 56.21 (1.4) & 99.14 (2.5) \\
\midrule

NSAE 
& $0\%$ & 99.99 (2.5) & 99.98 (2.5) & 00.09 (0.0) & 99.99 (2.5) & \textbf{77.03 (1.9)} & \textbf{99.90 (2.5)} & \textbf{97.02 (2.4)} & 99.58 (2.5) \\
& $20\%$ & 99.80 (2.5) & \textbf{99.99 (2.5)} & 04.82 (0.1) & 34.92 (0.9) & 36.04 (0.9) & 00.00 (0.0) & 05.16 (0.1) & \textbf{99.80 (2.5)} \\
& $30\%$ & \textbf{99.99 (2.5)} & 03.16 (0.1) & 00.32 (0.0) & \textbf{99.83 (2.5)} & 59.36 (1.5) & 15.90 (0.4) & 00.35 (0.0) & 99.69 (2.5) \\
& $50\%$ & 99.98 (2.5) & 34.36 (0.9) & \textbf{53.66 (1.3)} & 99.83 (2.5) & 12.36 (0.3) & 00.00 (0.0) & 19.33 (0.5) & 99.40 (2.5) \\
\midrule

CORAL 
& $0\%$ & 59.18 (1.5) & 90.61 (2.3) & 30.45 (0.8) & 00.80 (0.0) & 08.40 (0.2) & 55.06 (1.4) & 03.40 (0.1) & 69.94 (1.7) \\
& $20\%$ & 61.53 (1.5) & 12.64 (0.3) & \textbf{31.79 (0.8)} & \textbf{50.43} (1.3) & \textbf{33.74 (0.9)} & 41.31 (1.0) & \textbf{50.38} (1.3) & 67.54 (1.7) \\
& $30\%$ & 38.85 (1.0) & \textbf{99.99 (2.5)} & 0.00 (0.0) & 0.01 (0.0) & 22.96 (0.6) & 82.21 (2.1) & 0.00 (0.0) & 95.01 (2.4) \\
& $50\%$ & \textbf{99.99 (2.5)} & 38.85 (1.0) & 00.01 (0.0) & 00.00 (0.0) & 22.96 (0.6) & \textbf{82.21 (2.1)} & 00.01 (0.0) & \textbf{99.19 (2.5)} \\
\midrule

MTLS-RED 
& $0\%$ & \textbf{98.83 (2.2)} & {96.08 (2.2)} & {72.41 (2.1)} & \textbf{96.12 (2.4)} & 28.01 (0.7) & 80.99 (2.0) & 73.93 (1.8) & 73.23 (1.8) \\
& $20\%$ & 79.00 (2.0) & 78.85 (2.0) & 78.71 (2.0) & 78.68 (2.0) & 78.88 (2.0) & 79.13 (2.0) & 78.82 (2.0) & 79.01 (2.0) \\
& $30\%$ & 83.90 (2.1) & 80.70 (2.0) & 70.70 (2.0) & 83.80 (2.1) & 76.50 (1.9) & {81.96 (2.0)} & \textbf{85.00 (2.1)} & 77.10 (1.9) \\
& $50\%$ & 86.46 (2.0) & \textbf{89.33 (2.0)} & \textbf{79.20} (2.0) & 89.12 (2.0) & \textbf{78.99 (2.0)} & \textbf{89.33} (2.0) & 79.25 (2.0) & \textbf{79.64 (2.0)} \\
\bottomrule
\end{tabular}}
\end{table*}
This approach is inspired by the paper’s emphasis on the role of well-conditioned covariances for successful extrapolation, suggesting that diverse training correlations expand the set of feature configurations on which the model is trained, thus boosting generalization performance in truly novel test domains.
\begin{figure}[t]
    \centering
    \subfloat[BOTNET (0.0, 0.0; 0.6760)]{\includegraphics[width=0.24\textwidth,height=0.4\textheight,keepaspectratio]{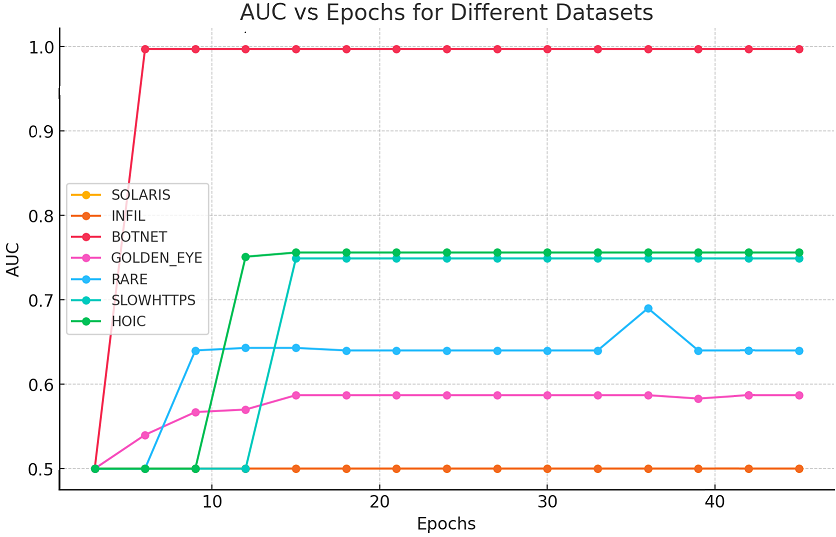}}\hfill
    \subfloat[BOTNET (0.5, 0.5; 0.7332)]{\includegraphics[width=0.24\textwidth,height=0.4\textheight,keepaspectratio]{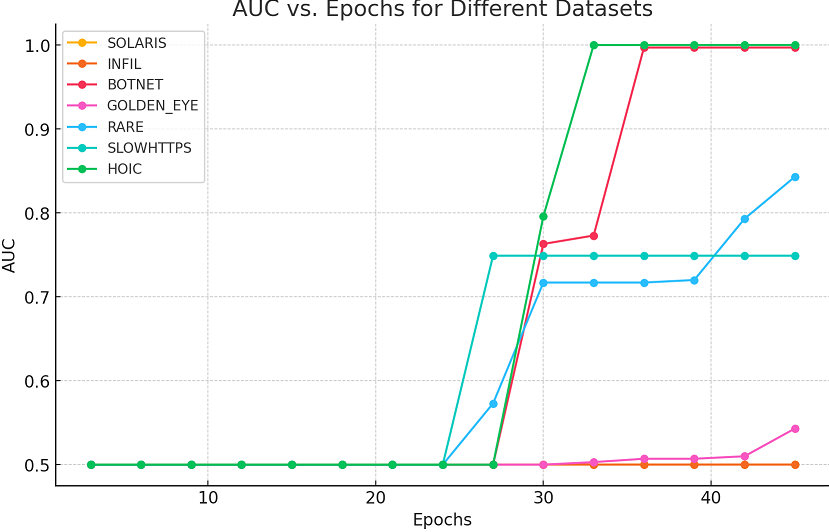}}\hfill
    \subfloat[BOTNET, SOLARIS (0.1,0.1; 0.7670)]{\includegraphics[width=0.24\textwidth,height=0.4\textheight,keepaspectratio]{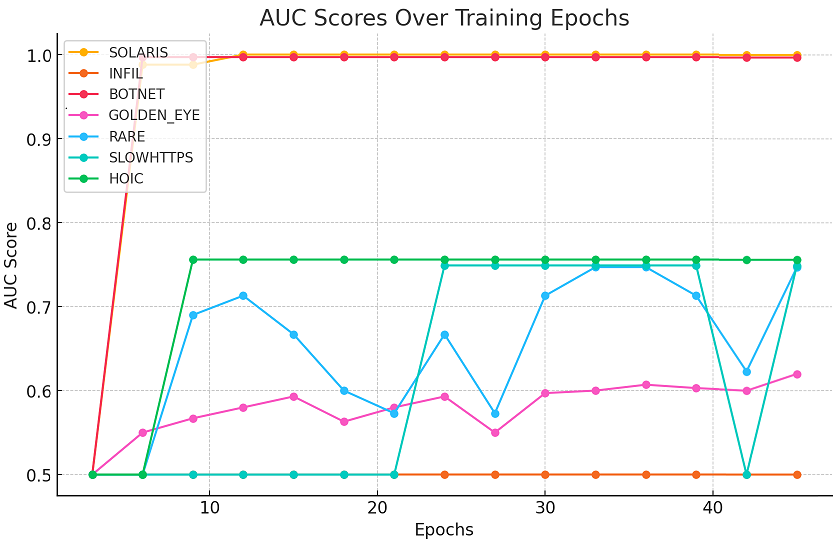}}\hfill
    \subfloat[ BOTNET, SOLARIS(0.01, 0.99; 0.7983)]{\includegraphics[width=0.24\textwidth,height=0.4\textheight,keepaspectratio]{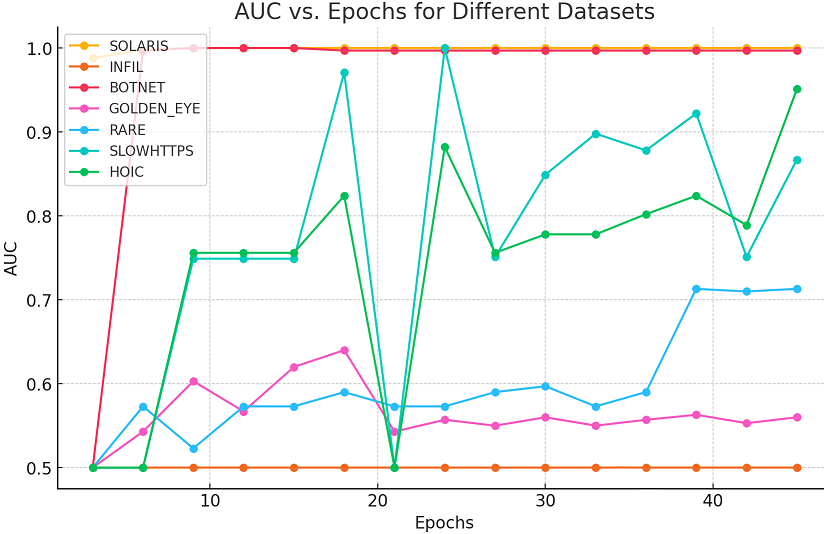}}

    %\vspace{3pt} % Adds space between rows to avoid overlap

   %  \subfloat[ INFIL (0.0, 0.0; 0.500)]{\includegraphics[width=0.24\textwidth,height=0.4\textheight,keepaspectratio]{FIGURES_AUC/mtl_red_auc_lrec_0_1_lmi__1_only_source.PNG}}\hfill
   %  \subfloat[SOLARIS (0.01, 0.99; 0.6073)]{\includegraphics[width=0.24\textwidth,height=0.4\textheight,keepaspectratio]{FIGURES_AUC/solaris_0_0.PNG}}\hfill
   %  \subfloat[SOLARIS, INFIL (0.01, 0.99; 0.590)]{\includegraphics[width=0.24\textwidth,height=0.4\textheight,keepaspectratio]{FIGURES_AUC/botnet_infil_0_0_0590.PNG}}\hfill
   % \subfloat[BOTNET, INFIL (0.01, 0.99; 0.6859)]{\includegraphics[width=0.24\textwidth,height=0.4\textheight,keepaspectratio]{FIGURES_AUC/infil_botnet_001_099_06859.PNG}}\hfill

    \vspace{3pt} % Adds space between rows to avoid overlap

    \subfloat[ GOLDEN-EYE (0.0, 0.0; 0.749)]{\includegraphics[width=0.24\textwidth,height=0.4\textheight,keepaspectratio]{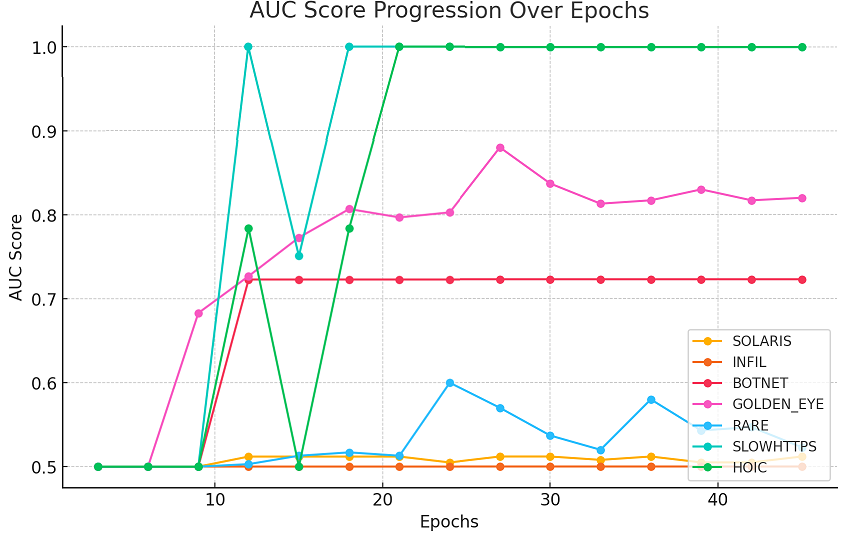}}\hfill
    \subfloat[ GOLDEN-EYE (0.01, 0.99; 0.7721)]{\includegraphics[width=0.24\textwidth,height=0.4\textheight,keepaspectratio]{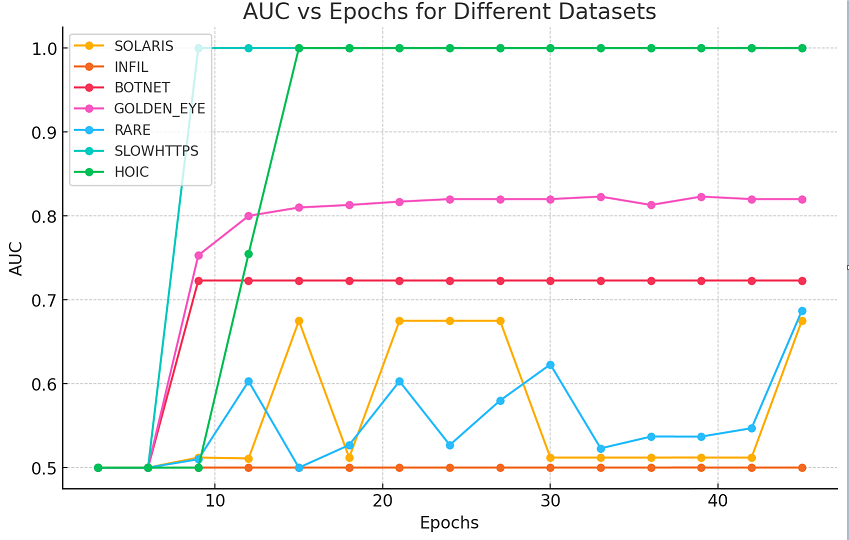}}\hfill
    \subfloat[ GOLDEN-EYE, BOTNET (0.1, 0.1; 0.849)
    ]{\includegraphics[width=0.24\textwidth,height=0.4\textheight,keepaspectratio]{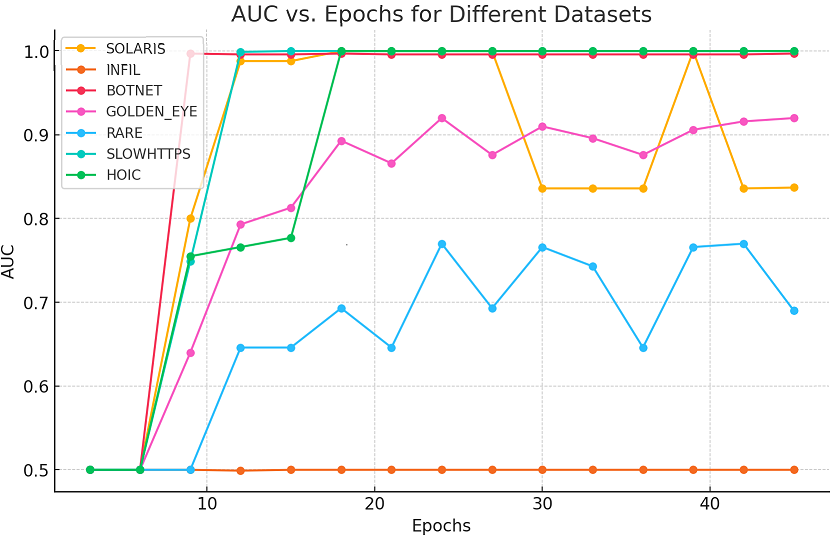}}\hfill
    \subfloat[ GOLDEN-EYE, BOTNET (0.1, 0.9; 0.8915)]{\includegraphics[width=0.24\textwidth,height=0.4\textheight,keepaspectratio]{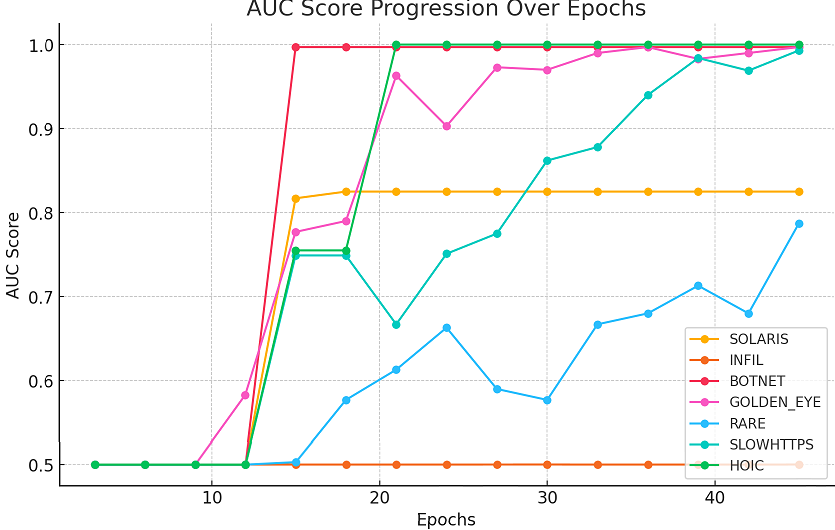}}
    \caption{From left to right, the plots show improved average AUC-ROC as datasets are combined and regularization is applied, enhancing generalization to unseen domains. We evaluate this using seven CIC-CSE-IDS attack datasets with equal benign samples, reporting results as (reconstruction weight, MI penalty, Average AUC on all datasets).}
    \label{ref:auc}
\end{figure}
DOS and DDOS share similar feature correlations, while MIRAI and WEB differ, and GOLDEN and SOLARIS exhibit distinct correlations from INFIL and BOTNET. We aim to enhance generalization by training on datasets with varying feature correlation structures while ensuring overlapping marginal distributions for effective extrapolation.
\footnote{\url{https://github.com/padmaksha18/MTRAE/blob/main/mtrae/mtl-reg-cse-cic-ids-V333333-noisy-equal-cross.ipynb}}

\begin{figure*}[htp]
    \centering

    % Top row: (a)-(e)
    \subfloat[]{\includegraphics[height=2.0cm,width=0.19\textwidth]{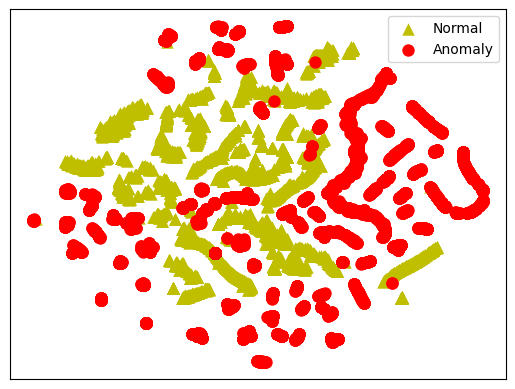}}%
    \subfloat[]{\includegraphics[height=2.0cm,width=0.19\textwidth]{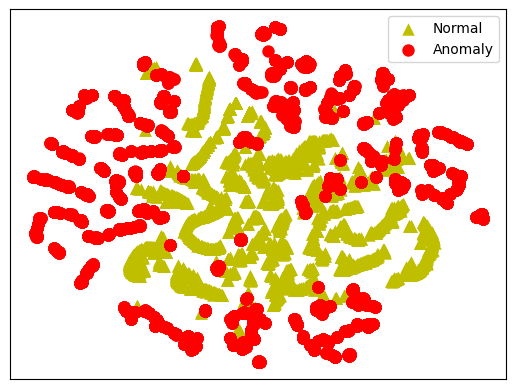}}%
    \subfloat[]{\includegraphics[height=2.0cm,width=0.19\textwidth]{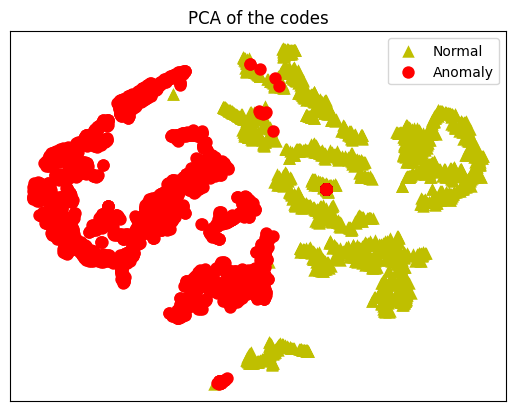}}%
    \subfloat[]{\includegraphics[height=2.0cm,width=0.19\textwidth]{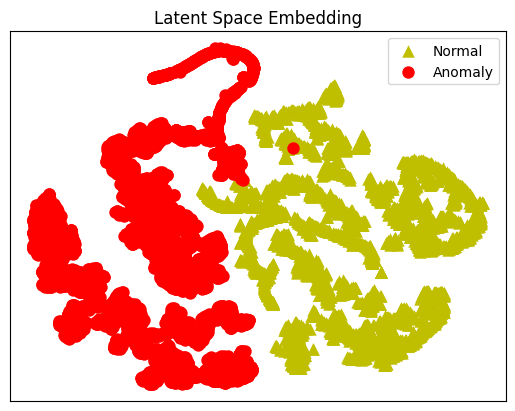}}%
    \subfloat[]{\includegraphics[height=2.0cm,width=0.19\textwidth]{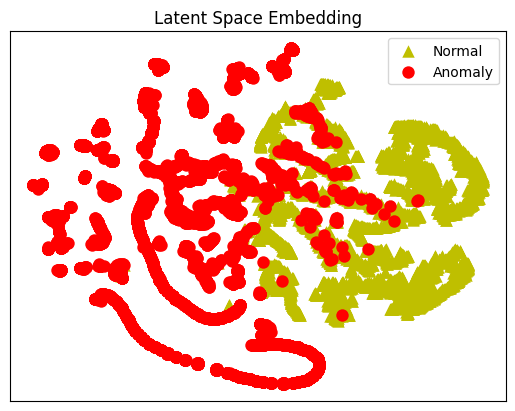}}%

    %\\[6pt]

    % Bottom row: (f)-(j)
    \subfloat[]{\includegraphics[height=2.0cm,width=0.19\textwidth]{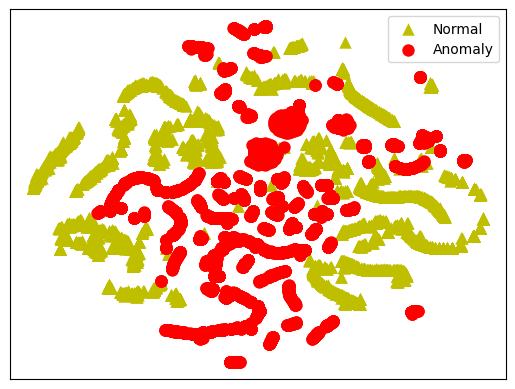}}%
    \subfloat[]{\includegraphics[height=2.0cm,width=0.19\textwidth]{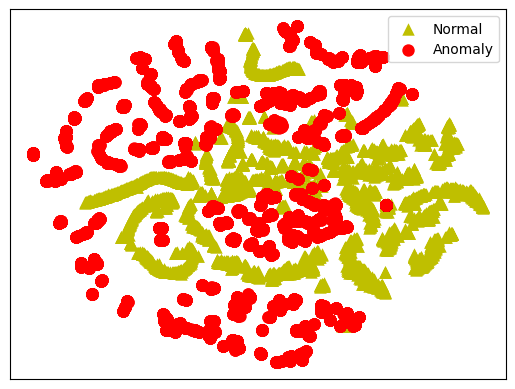}}%
    \subfloat[]{\includegraphics[height=2.0cm,width=0.19\textwidth]{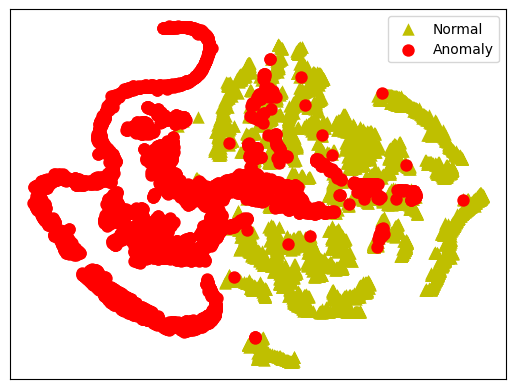}}%
    \subfloat[]{\includegraphics[height=2.0cm,width=0.19\textwidth]{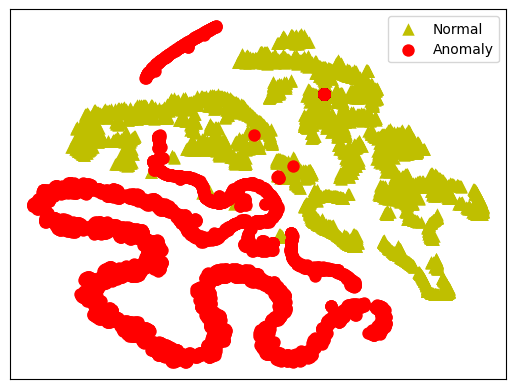}}%
    \subfloat[]{\includegraphics[height=2.0cm,width=0.19\textwidth]{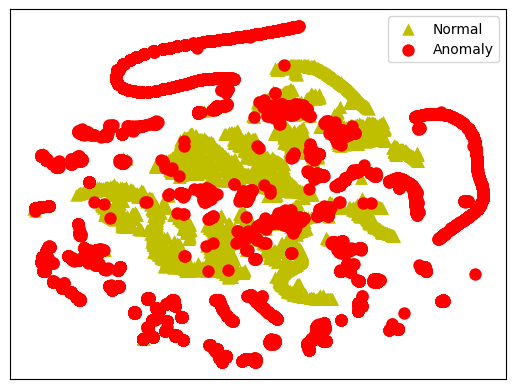}}%

    \caption{T-SNE projection of the latent space of without regularization case (bottom row) and MTL-RED (top row) for some of the attacks in CIC-IDS and CIC-IOMT/IOT: 
    SOLARIS, RARE, DOS, DDOS, and RECONAISSANCE. Subfigures (a)--(e) correspond to MTL-RED, and (f)--(j) to no regularization case.}
    \label{Fig:4}
\end{figure*}

\subsection{Hyperparameter Sensitivity} 
In the joint optimization framework, achieving an optimal balance between the compression regularization (\(\lambda\)) and the mutual information regularization (\(\beta\)) is crucial for ensuring strong generalization across all classes. The reconstruction loss, weighted by \(\lambda\), governs the degree of compression in the latent space—excessive compression may lead to the loss of essential features, while insufficient compression can result in overfitting to the input distribution. Meanwhile, the mutual information regularization, controlled by \(\beta\), acts as a de-correlation penalty, reducing redundant dependencies between the latent and input spaces. Properly tuning \(\beta\) ensures that the latent representation retains only the most discriminative information for classification. Our findings indicate that prioritizing entropy regularization (higher \(\beta\)) while reducing the emphasis on reconstruction loss (lower \(\lambda\)) yields the best overall model performance across diverse scenarios, reinforcing the importance of controlled compression and structured disentanglement in the latent space.
\subsubsection{Importance of kernel bandwidth}
In a non-parametric estimation method such as our mutual information penalty, the kernel bandwidth $\sigma$ plays a crucial role. By learning $\sigma$ jointly with the encoder and decoder, we adapt the kernel scale to match the data distribution’s complexity. We vary the proportion of cross-domain data in training, ranging from \text{0\%-50\%} of the source data, to analyze the effects of the de-correlation penalty and reconstruction regularization under different scenarios. In Tables \ref{Tab:1} and \ref{Tab:2}, we observe that as the proportion of cross-domain data in training increases, the performance of most baseline models deteriorates on the IN distribution. In particular, OOD domain data remain completely unseen throughout the training process. Table \ref{Tab:1} reveals an intriguing trend: as cross-domain data increases to $40\%-50\%$, adjusting the kernel bandwidth and hyperparameters $\beta$ and $\lambda$ allows us to train a model that achieves comprehensive generalization across all training and test datasets.
\begin{table*}[hbt!]
\centering
\renewcommand{\arraystretch}{1.2}
\setlength{\tabcolsep}{5pt}
\caption{We report \textbf{accuracy} (with std deviation) of the proposed and baseline methods on the CIC-IOMT/IOT dataset Other details are similar as Table 1.}
\label{Tab:2}
\resizebox{0.99\textwidth}{!}{
\begin{tabular}{lccccccccc}
\toprule
\textbf{Model} & \textbf{Percent} & \multicolumn{2}{c}{\textbf{SOURCE DOMAINS}} & \multicolumn{2}{c}{\textbf{CROSS DOMAINS}} & \multicolumn{4}{c}{\textbf{OOD DOMAINS}} \\
& & DDOS & DOS & RECON & SPOOF & MQTT & MIRAI & WEB & BENIGN \\
\midrule
MTAE & 0\% & 99.96 (2.5) & 99.99 (2.6) & 54.07 (1.5) & 46.98 (1.3) & 78.43 (2.1) & \textbf{99.97 (2.8)} & 30.55 (0.8) & 97.53 (2.8) \\
& 20\% & 99.99 (2.8) & 99.99 (2.7) & 98.66 (2.3) & 71.84 (1.4) & 78.43 (2.0) & 79.38 (1.8) & \textbf{30.55 (0.7)} & 97.53 (2.2) \\
& 30\% & 99.99 (2.4) & 99.99 (2.5) & \textbf{99.99 (2.5)} & \textbf{75.55 (1.9)} & \textbf{99.93 (2.9)} & 80.07 (2.3) & 21.38 (0.6) & 98.63 (2.3) \\
& 50\% & \textbf{99.99 (2.4)} & \textbf{99.99 (2.7)} & 98.53 (2.1) & 74.43 (2.2) & 89.42 (2.2) & 80.83 (2.2) & 29.02 (0.7) & 97.78 (2.7) \\
\midrule
MMD-AE & 0\% & 99.96 (3.0) & 99.96 (2.5) & 49.95 (1.3) & 41.32 (0.9) & \textbf{95.87 (2.7)} & 84.01 (1.8) & 21.15 (0.6) & \textbf{98.56 (2.9)} \\
& 20\% & \textbf{99.99 (2.8)} & \textbf{99.99 (2.7)} & 98.47 (2.9) & 70.54 (1.9) & 90.54 (2.7) & \textbf{84.01 (2.2)} & 42.53 (0.9) & 96.49 (2.1) \\
& 30\% & 99.99 (2.4) & 99.99 (2.8) & 99.11 (2.1) & 73.60 (1.7) & 89.42 (2.5) & 77.60 (2.2) & 58.04 (1.4) & 93.48 (2.4) \\
& 50\% & 99.61 (2.4) & 99.46 (2.4) & \textbf{99.30 (2.5)} & \textbf{78.23 (2.2)} & 94.84 (2.4) & 72.08 (2.1) & \textbf{67.21 (1.5)} & 92.41 (2.3) \\
\midrule
NSAE & 0\% & \textbf{99.99 (2.5)} & 99.99 (2.0) & 46.90 (1.1) & 32.42 (0.8) & 69.11 (1.7) & \textbf{99.90 (2.7)} & \textbf{56.05 (1.1)} & 94.68 (2.0) \\
& 20\% & 99.99 (3.0) & \textbf{99.99 (2.9)} & 97.77 (2.8) & 69.11 (1.9) & \textbf{99.90 (2.6)} & 99.87 (2.1) & 36.66 (0.9) & \textbf{97.71 (2.3)} \\
& 30\% & 99.99 (2.6) & 99.99 (2.9) & \textbf{98.70 (2.5)} & \textbf{71.29 (1.9)} & 99.71 (2.7) & 99.90 (2.1) & 43.83 (1.2) & 96.11 (2.0) \\
& 50\% & 99.99 (2.4) & 99.99 (2.7) & 98.62 (2.1) & 70.64 (2.0) & 99.89 (2.3) & 99.90 (2.6) & 49.70 (1.1) & 95.78 (2.4) \\
\midrule
CORAL & 0\% & \textbf{99.99 (2.6)} & 99.99 (3.0) & 98.53 (2.4) & 74.43 (2.2) & 89.42 (2.0) & \textbf{81.13 (2.3)} & 30.53 (0.8) & 98.73 (2.7) \\
& 20\% & 99.99 (2.2) & \textbf{99.99 (2.8)} & 98.53 (2.8) & 74.43 (2.2) & 89.42 (2.1) & 81.13 (2.1) & \textbf{42.30 (1.0)} & 98.73 (2.2) \\
& 30\% & 99.99 (2.4) & 99.99 (2.9) & \textbf{98.86 (2.5)} & 75.55 (1.9) & \textbf{99.90 (2.9)} & 80.08 (2.3) & 42.03 (0.9) & \textbf{98.73 (2.5)} \\
& 50\% & 99.99 (2.5) & 99.99 (2.7) & 99.11 (2.3) & \textbf{77.21 (2.0)} & 99.57 (2.7) & 80.51 (2.2) & 42.30 (1.0) & 98.73 (2.4) \\
\midrule
MTLS-RED & 0\% & 99.99 (2.7) & 99.99 (2.8) & 98.41 (2.4) & 78.92 (2.0) & 78.21 (1.9) & 76.04 (1.8) & 73.91 (1.7) & 87.67 (2.5) \\
& 20\% & 99.99 (2.6) & 99.99 (2.7) & 98.41 (2.3) & 78.92 (1.9) & 78.21 (1.8) & 99.75 (2.5) & 68.03 (1.6) & 91.83 (2.4) \\
& 30\% & 99.99 (2.8) & 99.99 (2.5) & 98.41 (2.8) & 78.21 (1.9) & 53.19 (1.5) & 99.87 (2.6) & 73.67 (1.8) & \textbf{93.05 (2.3)} \\
& 50\% & \textbf{99.99 (2.5)} & \textbf{99.99 (2.8)} & \textbf{98.98 (2.5)} & \textbf{81.17 (2.3)} & \textbf{95.87 (2.7)} & \textbf{99.88 (2.3)} & \textbf{75.91 (1.6)} & 91.20 (2.4) \\
\bottomrule
\end{tabular}}
\end{table*}
Figure \ref{ref:auc} illustrates the improvement in average AUC-ROC as datasets are combined and regularization is introduced. Both strategies—dataset combination and regularization—enhance generalization to unseen domains. The datasets are added strategically to improve coverage of unseen domains, ensuring a broader representation. In most cases, assigning \textit{higher weight to the MI penalty} while keeping the \textit{reconstruction weight minimal} leads to the best generalization performance. Figure \ref{Fig:precision_recall_auc_rare} demonstrates the impact of the regularization penalty on the RARE dataset. Our results indicate that incorporating regularization—and particularly increasing the weight on the mutual information (decorrelation) penalty—leads to improved and more stable precision, recall, and AUC when training on the combined GOLDEN-EYE and SOLARIS dataset.

\begin{table*}[hbt!]
\centering
\renewcommand{\arraystretch}{1.3}  % Adjust row spacing
\setlength{\tabcolsep}{6pt}  % Adjust column spacing
\caption{We report \textbf{accuracy} (with std deviation) of the proposed and baseline methods on the Arrythmia dataset. For each case, the dataset contains equal amount of normal and anomaly samples.}
\label{Tab:3}
\resizebox{0.85\textwidth}{!}{
\begin{tabular}{lcccccc}
\toprule
\textbf{Model} & \textbf{Percent (\%)} & \multicolumn{2}{c}{\textbf{SOURCE DOMAINS}} & \multicolumn{2}{c}{\textbf{CROSS DOMAINS}} & \textbf{OOD DOMAINS}  \\
 & & VEB & BENIGN & SVEB & Q & F \\
\midrule

MTAE 
& $50\%$ & 99.17 (2.48) & 62.94 (1.57) & 60.53 (1.51) & 90.41 (2.26) & {60.00 (2.50)} \\

\midrule

MMD-AE 
& $50\%$ & 98.58 (2.46) & 63.63 (1.59) & 44.90 (1.12) & 93.33 (2.33) & 89.66 (2.24)  \\

\midrule

NSAE 
& $50\%$ & 97.18 (2.43) & 72.71 (1.82) & 41.70 (1.04) & 80.00 (2.00) & 90.17 (2.25)  \\

\midrule

CORAL 
& $50\%$ & 97.87 (2.45) & 58.24 (1.46) & 69.70 (1.74) & 95.26 (2.38) & 73.33 (1.83)  \\

\midrule

MTLS-RED 
& $50\%$ & \textbf{99.34 (2.48)} & \textbf{77.49 (1.94)} & \textbf{73.48 (1.84)} & \textbf{95.52 (2.39)} & \textbf{93.33} (2.33)  \\

\bottomrule
\end{tabular}}
\end{table*}

In Figure~\ref{Fig:4}, we visualize the latent space representations of a standard multi-task encoder-decoder model without regularization and our proposed model incorporating the MI penalty. We observe improved clustering of source, cross-domain, and target domain classes when regularization is applied, demonstrating its effectiveness in structuring the latent space. As shown in Table~\ref{Tab:2}, increasing the proportion of cross-domain data during training significantly enhances classification performance across OOD datasets, such as WEB and SPOOFING attacks. Likewise, Table~\ref{Tab:3} presents the evaluation of our method on the Arrhythmia dataset, where the model is trained on the normal and VEB classes while considering all other anomaly classes—SVEB, Q as cross-domain, and F—as OOD test class. In Table~\ref{Tab:EMG}, we evaluate our approach on a time-series dataset (EMG Gesture Recognition)  and compare with the baselines.

\begin{table}[]
\centering
\setlength{\tabcolsep}{1pt}
\small
\label{tab:results}
        \begin{tabular}{c|c|c|c|c}
        MODEL & Domain1 & Domain2  & Domain3 & Domain4
        \\ \hline
        DIFEX & 65.02 $\pm$ 2.00 & 66.15 $\pm$ 2.50 & $\mathbf{64.06}$ $\pm$ $\mathbf{2.00}$ & 62.98 $\pm$ 2.00 
        \\ 
        CORAL & 52.39 $\pm$ 2.00 & 52.51 $\pm$ 2.50 & 53.89 $\pm$ 2.00 & 57.06 $\pm$ 2.00 
        \\ 
        MTL-RED & $\mathbf{66.41}$ $\pm$ $\mathbf{3.40}$ & $\mathbf{66.30}$ $\pm$ $\mathbf{2.50}$ & 55.92 $\pm$ 2.00  & $\mathbf{65.82}$ $\pm$ $\mathbf{2.50}$ 
        \\
        \hline
        \end{tabular}
        \normalsize
\caption{Performance (accuracy \%) of MTL-RED, DIFEX, CORAL with EMG time series dataset divided into 4 domains each consisting of 6 classes for all the 9 persons. }
\label{Tab:EMG}
\end{table}
\section{Conclusion} Our paper addresses the challenge of detecting novel and out-of-distribution (OOD) anomalies through domain generalization techniques. By training on multiple source and cross-domain datasets with distinct correlation structures, we aim to increase the coverage to generalize to unseen anomaly classes. Subsequently, guided by the principle of relevant information preservation (PRI), our regularization steers the cross-entropy loss in latent space to retain essential features to achieve domain generalization. Real-world cybersecurity and healthcare datasets often exhibit different correlation patterns (varying joint distributions) among the different classes, which can be exploited to increase the coverage for extrapolation to new, unseen domains. Future work will further explore methods for latent space anomaly detection.

\section*{Acknowledgment} We gratefully acknowledge the support of the Virginia Tech National Security Institute (VTNSI) and the Deloitte \& Touche LLP, USA, for supporting this research. We also extend our sincere thanks to our collaborators at Deloitte — Ajay Kumar, Alison Hu, Sanmitra Bhattacharya, and Edward Bowen for their insightful contributions.

%
% ---- Bibliography ----
%
% BibTeX users should specify bibliography style 'splncs04'.
% References will then be sorted and formatted in the correct style.
%
\bibliographystyle{splncs04}
%\bibliography{mybibliography}

%% Note that this preceding line implies that you store your BibTeX references in a file called 'mybibliography.bib'. If you instead store your references in a file with a different name, for instance 'references.bib', the preceding line should read '\bibliography{references}'. Whatever you do, DO NOT put the file name extension .bib inside the \bibliography command; this will trip up LaTeX compilers. 
%
% If you do not want to use BibTeX, you can also type up the bibliography exactly as you see fit, using the following structure:

%\section{Appendix}
\section{Joint Optimization and the Principle of Relevant Information Preservation (PRI)}

Our approach is motivated by Tishby's Principle of Relevant Information (PRI), which states that an optimal representation should preserve only the information in the input that is necessary for the task at hand, while discarding irrelevant details. In our context, the goal is to learn a latent representation \( Z \) from the input \( X \) that is both predictive of the target \( Y \) and minimally influenced by spurious correlations present in \( X \). To achieve this, we jointly optimize a loss function that combines three key components:
\begin{enumerate}
    \item \textbf{Cross-Entropy Loss} (\(\mathcal{L}_{\text{CE}}\)): This term ensures that the latent representation \( Z \) is discriminative enough to accurately predict the target \( Y \).
    \item \textbf{Reconstruction Loss} (\(\mathcal{L}_{\text{recon}}\)): This term (e.g., mean squared error) forces \( Z \) to retain sufficient information to reconstruct the input \( X \), thereby preventing excessive compression.
    \item \textbf{Mutual Information Penalty} (\(\mathcal{L}_{\text{MI}}\)): By penalizing the mutual information between \( X \) and \( Z \), this term encourages the model to discard spurious and domain-specific correlations, leading to a more invariant and disentangled latent space.
\end{enumerate}

The overall objective can be written as:
\begin{equation}
    \mathcal{L} = \mathcal{L}_{\text{CE}} + \lambda_{\text{recon}} \, \mathcal{L}_{\text{recon}} + \lambda_{\text{MI}} \, \mathcal{L}_{\text{MI}},
    \label{eq:joint_loss}
\end{equation}
where \(\lambda_{\text{recon}}\) and \(\lambda_{\text{MI}}\) are hyperparameters that balance the trade-off between reconstruction fidelity and the strength of the decorrelation (compression) penalty.

In the framework of PRI (proposed by Tishby and later on implemented in various contexts), we aim to minimize the mutual information between \( X \) and \( Z \) while maintaining high mutual information between \( Z \) and \( Y \). This idea is often expressed as:
\begin{equation}
    \mathcal{L}_{\text{PRI}} = I(X;Z) - \beta \, I(Y;Z),
    \label{eq:pri}
\end{equation}
where:
\begin{itemize}
    \item \( I(X; Z) \) quantifies the total information that the latent representation \( Z \) retains about \( X \).
    \item \( I(Y; Z) \) measures the information in \( Z \) that is useful for predicting \( Y \).
    \item \(\beta\) is a parameter controlling the trade-off between compression (minimizing \(I(X;Z)\)) and predictive power (maximizing \(I(Y;Z)\)).
\end{itemize}

In practice, our joint loss in Equation~\eqref{eq:joint_loss} serves as a proxy for the PRI objective in Equation~\eqref{eq:pri}:
\begin{itemize}
    \item The \(\mathcal{L}_{\text{CE}}\) term drives \( Z \) to retain information relevant to \( Y \) (i.e., maximizing \(I(Y;Z)\)).
    \item The combination of \(\mathcal{L}_{\text{recon}}\) and \(\mathcal{L}_{\text{MI}}\) encourages \( Z \) to compress \( X \) by preserving only the necessary information and discarding spurious correlations, effectively minimizing \(I(X;Z)\).
\end{itemize}

Moreover, our implementation of the mutual information penalty is based on the Renyi entropy of kernel matrices computed from \( X \) and \( Z \), with kernel bandwidths \( s_x \) and \( s_y \) that are adjusted during training. This enables the model to learn an optimal level of decorrelation, ensuring that the latent space does not overfit to domain-specific artifacts while still preserving the relevant structure needed for accurate classification and reconstruction. In summary, our joint optimization framework, which integrates classification, reconstruction, and decorrelation, is a practical instantiation of the PRI principle. By carefully balancing these objectives, our model is guided to learn a compressed yet task-relevant latent representation that generalizes effectively across domains.

\end{document}